
%
%
\input jnl/jnl.tex
\input jnl/reforder.tex
\input jnl/eqnorder.tex
\title
Fracton Superconductivity in
Quasicrystals --- A Theoretical Possibility
Suggested by  Experiment$^*$
\author  K. Moulopoulos  and F. Cyrot-Lackmann
\affil  Laboratoire d' Etudes des Propri\'et\'es Electronique
des Solides, CNRS,  38042 Grenoble
\abstract
Scattering of electrons due to fractons
can result in a resistivity that {\it decreases} with temperature.
Such a behavior
also appears in real quasicrystals.
If this is then attributed to fracton scattering,
fracton-superconductivity would be theoretically possible.
By fitting with the scattering result the experimental resistivity data
for the AlPdMn quasicrystal, we estimate the
corresponding $T_c$.
This effective fracton interpretation, not unexpected for a self-similar
system,
is also found consistent with other
experiments on thermal and acoustic properties of the AlPdMn quasicryst not
offal.

$$ $$
PACS numbers: 61.44.+p, 74.20.-z, 61.43.Hv, 72.10.-d
$$ $$
$$ $$
$$ $$
$^*$ Submitted to Phys. Rev. Letters (April 1995)

\endtopmatter

The quasicrystalline state\refto{janot} is a new and  novel state of matter
discovered
 a decade ago\refto{gratias}, with anomalous physical properties
 that  still pose a challenge to a fundamental
theoretical understanding. Most concepts of conventional Solid State Physics
are inappropriate for a rigorous theoretical modeling of ideal quasicrystalline
 systems. The basic reason is  the  lack of periodicity, that makes concepts
from
Bloch theory invalid, but also the lack of randomness, in its usual sense,
that, rigorously speaking,  prevents one
from using concepts pertaining to disordered systems.
In fact, from a structural point of view,
 quasicrystals can be viewed as being in between periodic
and purely disordered systems. Their seemingly irregular structure incorporates
a complete
(aperiodic) long-range order.
Nevertheless, there have been in the past decade important theoretical
developements
towards a complete structural characterization, especially based on the
hyperspace construction or, alternatively, on matching rules in physical space.
However, {\it given} the structure of an ideal quasicrystal, almost nothing is
known with certainty concerning its electronic properties.
It is only in 1-d quasiperiodic chains that rigorous results are
known\refto{1d},
and this is basically a consequence of the trivial topology.
Most of these results
 manifest a new type of electronic state that has been called
``critical", which can  also be viewed as having a character that is between
extended and
localized states (the former being the generic electronic state in an ideally
periodic system, the latter being the one in a sufficiently disordered system).
What is fascinating
(although not entirely unexpected) is that these critical states have
a {\it multifractal} nature, basically because of a type of {\it
self-similarity}
that characterizes a typical quasiperiodic structure and that is
further discussed
below.

Experimentally there has been active research on structural
aspects\refto{structure} with few questions still being unresolved,
concerning especially the actual positions of atoms.
In the past few years there has also been intense activity
 on electronic
properties\refto{electronic} as well.
The latter demonstrate a number of {\it anomalous} transport properties
that seem to be opposite to the typical behavior of metallic materials.
The simplest of these, namely the temperature dependence
of the resistivity, is the focus of our work below, and will be given a new
interpretation, related to the fractal nature of the quasicrystalline state.

 It is known that in  quasicrystalline phases there exist atomic
clusters with hierarchical and quasiperiodic packings\refto{janot2}, the
existence of which
is ultimately related to an already mentioned type of self-similarity
 of the quasicrystalline
state; this is  scale-invariance
 under appropriate inflation/deflation transformations.
It may therefore be  expected that
quasicrystals may effectively have  properties that
 generically appear in fractal systems. In particular,
the vibrational properties may be describable in terms of
effective {\it fracton} spectra (quanta of special localized modes
in fractal systems), a  concept
introduced\refto{fracton}  shortly before the discovery of quasicrystals
and used in approximate theoretical modeling of certain disordered
systems\refto{rmp}.
We want to emphasize in this work the
interesting fact that such an
assumption of fractal  vibrational behavior, and in particular the
simulation of the vibrating
quasicrystalline structure with a vibrating  bond-percolation
network\refto{percolation}
with the proper  exponents, can account for the anomalous
transport properties of this system. In more detail, a simple
electron-scattering calculation due to fractons associated with these
vibrations, can lead to a resistivity that $\it{decreases}$ with temperature,
a behavior opposite to the normal electron-phonon scattering behavior
manifested in metallic materials.
On the other hand, it was already mentioned that
quasicrystals also behave anomalously; the experimentally
measured resistivity\refto{berger}, for example, always shows the same
qualitative behavior, namely a   resistivity {\it lowering} with temperature.
  The basic proposition in the present Letter is to relate the two
qualitatively similar  results
in a quantitative  manner. Hence,
by  attributing  the experimental behavior
of the resistivity of real quasicrystalline materials to an effective
fracton spectrum (and to the eventual electron-fracton scattering),
 we determine the
corresponding electron-fracton coupling. On the other hand,
 the mere existence of fractons and the scattering of electrons off them,
 will in principle lead to an attractive electron-electron interaction
(in a way similar to the conventional Fr\"ochlich\refto{frochlich}
 electron-phonon mechanism)
and will in principle result in superconductivity due to the
corresponding fracton-mediated electron-pairing.

In this Letter: 1) we use the above mentioned electron-fracton
scattering result to
determine the electron-fracton coupling from the
temperature-dependent resistivity of the AlPdMn quasicrystal,
2) we show evidence that this effective fracton interpretation is consistent
with other experiments on the same material,
 namely,  measurements of specific heat,
thermal conductivity, and temperature-dependent velocity of sound, and
3) we use
the above
 determined electron-fracton
coupling  within an approximate form of
the Eliashberg theory\refto{eliashberg} of strong-coupling
superconductivity, to
give an estimate of the range of the critical temperature
$T_c$ of the corresponding fracton-mediated
superconductivity.

In more detail, a  calculation\refto{costas} to second order in perturbation
theory
with respect to electron-fracton coupling, leads to a scattering time
that is approximately given by
$${1 \over {\tau(E_F)}} = (A\; {{n e^2}\over m})  \Biggl(
\ln{ \bigl({{\hbar \omega_F}\over {k_B T-\hbar \omega_c}} \bigr)
} \Biggr)^2 \eqno(tau)$$
with
$$(A\; {{n e^2}\over m})={{2 \pi}\over \hbar} N(E_F)
\Biggl(
{{\tilde{d} N_i 4 \lambda_0^2} \over
{2 \rho \xi^2 \omega_F^{\tilde{d}}\; \omega_c^{2 \tilde{d}/D} }}
\Biggr)^2 \eqno(A) $$
provided that one uses a fractal dimension $D=4$ and a fracton dimension
$\tilde{d}=4/3$, which are the proper values
for a bond-percolation network at the marginal
dimensionality $d=6$ where both mean-field exponents and hyperscaling
are valid\refto{stauffer}.
In \(tau) and \(A) $\rho$ is the bulk mass-density, $N_i$ is the total number
of atoms, $\lambda_0$ is the electron-fracton coupling constant,
$\xi$ is a characteristic length of fractal behavior (see below and equ. (8)),
$\omega_F$ is the fracton analog of the
Debye frequency, $\omega_c$ is the phonon-fracton
crossover frequency, and $N(E_F)$ is the electronic density of states per spin
at the Fermi level.
Then, a simple scattering-time approximation for the resistivity
(equivalent to a neglect of the usual  $\cos({\vec k. \vec k^\prime})$ factor
in the Boltzmann equation) is expected to be  a good approximation
in quasicrystals\refto{ash}
 because of the
stochastically fragmented $\vec k$-space.
As a result, the resistivity is approximately given by
$$\rho = A \;
\Biggl(
\ln{ \bigl({{\hbar \omega_F}\over {k_B T-\hbar \omega_c}} \bigr)} \Biggr)^2
\eqno(resistivity)$$
for $\hbar \omega_c  <  k_B T  << \hbar \omega_F$, and is seen to
$\it{decline}$
with temperature.

We now discuss the possibility of superconductivity caused by the above
electron-fracton scattering, which can in principle serve as a mechanism for
  electron-pairing.
By using the corresponding fracton density of states
$$ N(\omega)={{\tilde{d} N_i \omega^{\tilde{d}-1} }
\over {\omega^{\tilde{d}}_F}}  \eqno(density)$$
 and fracton dispersion relation
$$ \omega = \omega_c \bigl( {\xi \over 2} q \bigr)^{D/\tilde{d}} \eqno(omega)$$
in the framework of the
Eliashberg theory of strong-coupling superconductivity\refto{eliashberg}
 we arrive at an Eliashberg function
$$ \alpha^2 F(\omega) =
{{N(E_F) \tilde{d} N_i 4 \lambda_0^2}
\over
{\omega_F^{\tilde{d}}\; 2 \rho \xi^2\; \omega_c^{2 \tilde{d}/D}}}.
\eqno(eliashberg)$$
The  corresponding superconducting coupling paramemeter,
defined in the standard way by
   $\;\lambda (T=0) = 2
\int_{\omega_c}^{\omega_F} {\alpha^2 F(\omega)  {{d \omega} \over \omega} },
\;$ is then found to be  of the form
$$\lambda (T=0) =
{{2 N(E_F) \tilde{d} N_i 4 \lambda_0^2}
\over
{\omega_F^{\tilde{d}}\; 2 \rho \xi^2\; \omega_c^{2 \tilde{d}/D}}}\;
\ln{\bigl({\omega_F \over \omega_c}\bigr)}. \eqno(lambda) $$

We now need to estimate the electron-fracton coupling $\lambda_0^2$
from experimental measurements of resistivity in combination with
\(resistivity).
By fitting the resistivity data\refto{berger} on
the quasicrystal $Al_{70.5}Pd_{22}Mn_{7.5}$ with
 the expression \(resistivity),
for $100 K \leq T  \leq  150 K$, we obtain
$\hbar \omega_c \sim 0.4 \;K,$
$\hbar \omega_F \sim 19000 \;K,$
and
$A \sim 300 \; \mu \Omega cm$.

We should point out here that the very low value for
the crossover frequency $\omega_c$ is
actually expected due to the wide plateau measured recently\refto{chernikov}
in the thermal conductivity of the AlPdMn system. This is related to a
large characteristic length $\xi$ below which we effectively have fractal
behavior (or, alternatively,  above which the system looks homogeneous).
In the case of the AlPdMn system, and from the relation
$$ {\omega_F \over \omega_c}=({\xi \over a_0})^{D/\tilde{d}} \eqno(omegaF) $$
we find $\xi \simeq 36 a_0$, where $a_0$ is the lowest atomic length scale,
which we here take as the Bohr radius.

We now show that the above is also consistent with other experiments
on the same material:
Measurements of low-temperature specific heat\refto{specificheat} can be
 rather well reproduced through a phonon-fracton crossover interpretation.
Indeed, following the calculation of Avogadro {\it et al.}\refto{avogadro} we
arrive at
results in reasonable agreement with experimental points shown in
Fig. 1 of ref.17. In addition,
the rise of the thermal conductivity ($\kappa$)
 above the plateau\refto{chernikov}
seems to be consistent with the linear decrease of the
relative variation (${\delta v_s \over v_s}$) of the velocity of
sound ($v_s$)  with temperature\refto{zarembowich}
 (what could be called the ``Bellessa effect"\refto{bellessa})  if this
rise is explained through a theory of  phonon-assisted fracton-hopping
proposed by Orbach {\it et al.}\refto{orbach}.
In this theory, the two slopes
are related by
$$\Biggl({{ {\delta v_s \over v_s} }\over T} \Biggr) =
-{0.1 \over k_B} \Bigl( {\xi^2 \over {2 \pi^2 v_s}} \Bigr)
\Bigl({\kappa \over T} \Bigr). \eqno(slope) $$
The experimentally measured value\refto{zarembowich} of
$\Biggl({{ {\delta v_s \over v_s} }\over {T}} \Biggr) \simeq
 - 1.818 \times{{10^{-5}}\over {K}}$ (determined
 from the linear decrease
in the temperature range $5\; K < T < 27\; K$ and
 at the lowest measured
frequency of $\omega \sim 28\; MHz$)
then yields through \(slope)
a value of
$\Bigl({\kappa \over T} \Bigr) \simeq 8.889 \times  10^{-5} \;{{W}\over {K^2
m}}$,
 which is in turn reasonably consistent with the experimental data above the
plateau
shown in  Fig.2 of ref. 16.

By taking, therefore, the above fracton interpretation at face value,
we make use of the above parameters with an effective density  $r_s \sim 6$
and an effective electron-to-atom ratio $Z \simeq 1.5$,
 and together with  some uncertainty
associated with the effective electronic density of states
(we use a $N(E_F)$ being between ${1 \over 10}$ and ${1 \over 5}$
of the free-electron density of states), we arrive at the
following
estimate for  the range of  $\lambda$ values
$$ 1.26  <  \lambda (T=0)   <   1.79 \eqno(lambdafinal)  $$
Finally, we use  Kresin's expression\refto{kresin} for $T_c$
$$T_c = {{0.25  <\omega^2>^{1/2}} \over
{\sqrt{\exp{(2/\lambda_{eff})}-1}}} \eqno(Tc)$$
which is supposed to be valid for any arbitrary strength $\lambda$.
The parameter $\lambda_{eff}$ is defined by
$$\lambda_{eff}={{\lambda-\mu^*}\over
{1+2 \mu^*+\lambda \mu^* t(\lambda)}}$$
and
$$t(\lambda) \simeq 1.5 \exp{ \bigl(-0.28 \lambda \bigr)}.$$
The characteristic frequency appearing in \(Tc) is defined by
$$<\omega^2>={{
\int_{\omega_c}^{\omega_F} {\alpha^2 F(\omega) \omega d \omega}
} \over {
\int_{\omega_c}^{\omega_F} {\alpha^2 F(\omega)  {{d \omega} \over \omega} }
}} $$
which, with use of \(eliashberg), leads to $ <\omega^2>^{1/2} \simeq 4000 \;
K.$
We also need an estimate for the Coulomb parameter $\mu^*$ which we take from
Thomas-Fermi screening theory. This parameter is defined by
$$\mu^* = { \mu \over {1+ \mu \ln{({\omega_p \over \omega_F})}}}$$
with $\mu$ given by a Fermi surface average, namely
$$\mu=N(E_F)  < {{4 \pi e^2}\over {V (k^2+k_0^2)}} >_{_{_{FS}}}\;\;\;
\simeq {k_0^2 \over {8 k_F^2}} \ln{ \Bigl(4 {k_F^2 \over k_0^2} +1 \Bigr)}$$
with the Thomas-Fermi screening wavevector $k_0$ given by
$${k_0^2 \over k_F^2}=({16 \over {3 \pi^2}})^{2/3} \; r_s $$
and the plasma frequency given by
$$\omega_p =4.71 \times  1.16 \times  10^4/r_s^{3/2} \;K. $$
For an effective $r_s \sim 6$ this yields $\mu^*=0.816$, which signifies
a strong electron-electron interaction, that is actually expected
for quasicrystals from interpretation of transport data\refto{klein1,klein2}.
Substituting all this into \(Tc)
 with an effective $r_s \sim 6,$
we obtain the following estimate of the range of critical temperatures
$$ 0.23 \;\; K \;\; < \;\; T_c \;\; <\;\; 17.2 \;\;  K. \eqno(Tcfinal) $$
(Note that
the critical temperature can actually be much higher, for smaller values
of $\mu^*$, that can result, for example, from a
model different from the Thomas-Fermi theory).

Although this type of superconductivity is in principle possible,
it is expected that there are other physical mechanisms
in AlPdMn that might
prohibit its appearance.
The existence, for example, of magnetic moments, possible spin-orbit
couplings etc. may have such a destructive effect and render this
fracton-superconductivity unobservable.
The question then of how to experimentally overcome these prohibiting  factors,
in this or some other quasicrystalline material,
becomes of interest and would deserve further investigation.

 M. Cyrot and N.W. Ashcroft  are acknowledged for helpful discussions.
This work has been supported by  the European Union within the
Human Capital and Mobility Program.
\references

\refis{gratias}
D. Shechtman, I. Blech, D. Gratias and J.W. Cahn,
Phys.Rev.Lett. {\bf 53}, 1951 (1984)

\refis{janot}
C. Janot, ``Quasicrystals: A Primer" (Oxford Science Publications, 2nd ed,
1994)

\refis{structure}
For a summary of experimental activity in this field see
``Lectures On Quasicrystals", ed. F. Hippert and D. Gratias (Les Editions
de Physique Les Ulis, 1994)

\refis{electronic}
For a rather complete review see C. Berger, ref. 3

\refis{fracton}
S. Alexander and R. Orbach, J. Phys. (Paris) Lett. {\bf 43}, L-625 (1982)

\refis{percolation}
R. Orbach, Science {\bf 231}, 814 (1986)

\refis{berger}
P. Lanco, T. Klein, C. Berger, F. Cyrot-Lackmann, G. Fourcaudot and
A. Sulpice, Europh. Lett. {\bf 18}, 227 (1992)

\refis{frochlich}
H. Fr\"ohlich, Phys. Rev. {\bf 79}, 845 (1950)

\refis{costas}
K. Moulopoulos, to be published

\refis{stauffer}
D. Stauffer, Phys. Rep. {\bf 54}, 1 (1979)

\refis{ash}
N. W. Ashcroft, personal communication

\refis{eliashberg}
G.M. Eliashberg, Zh. Eksp. Teor. Fiz. {\bf  38}, 966 (1960)
(Sov. Phys. JETP {\bf 11}, 696 (1960))

\refis{chernikov}
M.A. Chernikov {\it et al.}, Phys. Rev. B {\bf 51}, 153 (1995)

\refis{specificheat}
M.A. Chernikov {\it et al.}, Phys. Rev. B {\bf 48}, 3058 (1993)

\refis{avogadro}
A. Avogadro {\it et al.} Phys. Rev. B {\bf  33}, 5637 (1986)

\refis{zarembowich}
N. Vernier, G. Bellessa, B. Perrin, A. Zarembowitch and M. De Boissieu,
Europh. Lett. {\bf 22}, 187 (1993)

\refis{bellessa}
G. Bellessa, Phys. Rev. Lett. {\bf  40}, 1456 (1978)

\refis{orbach}
A. Jagannathan, R. Orbach and O. Entin-Wohlman, Phys. Rev. B {\bf 39},
13465 (1989)

\refis{kresin}
V.Z. Kresin, Phys. Lett. A {\bf 122}, 434 (1987)

\refis{1d}
M. Kohmoto, L. P. Kadanoff, and Ch. Tang,  Phys. Rev. Lett.
{\bf  50}, 1870  (1983); S.
 Ostlund, R. Pandit, D. Rand, H. J. Schellnhuber,
and E. D. Siggia, {\it ibid.} {\bf 50}, 1873  (1983);
For rigorous theorems on more general quasiperiodic chains
see J. Bellissard, A. Bovier, and J. M. Ghez,
Comm. Math. Phys. {\bf 135}, 379 (1991)
and
J. M. Luck, Phys. Rev. B {\bf 39}, 5834 (1989)

\refis{janot2}
C. Janot and M. de Boissieu, Phys. Rev. Lett. {\bf  72}, 1674 (1994)

\refis{rmp}
T. Nakayama, K. Yakubo, and R. Orbach, Rev. Mod. Phys. {\bf 66}, 381 (1994)

\refis{klein1}
T. Klein, H. Rakoto, C. Berger, G. Fourcaudot, and F. Cyrot-Lackmann,
Phys. Rev. B {\bf 45}, 2046 (1992)

\refis{klein2}
T. Klein, C. Berger, G. Fourcaudot, J. C. Grieco, P. Lanco, and
F. Cyrot-Lackmann, Journ. Non-Cryst. Sol. {\bf 156-158}, 901 (1993)

\endreferences

\end

\\
\catcode`@=11
\newcount\tagnumber\tagnumber=0

\immediate\newwrite\eqnfile
\newif\if@qnfile\@qnfilefalse
\def\write@qn#1{}
\def\writenew@qn#1{}
\def\w@rnwrite#1{\write@qn{#1}\message{#1}}
\def\@rrwrite#1{\write@qn{#1}\errmessage{#1}}

\def\taghead#1{\gdef\t@ghead{#1}\global\tagnumber=0}
\def\t@ghead{}

\expandafter\def\csname @qnnum-3\endcsname
  {{\t@ghead\advance\tagnumber by -3\relax\number\tagnumber}}
\expandafter\def\csname @qnnum-2\endcsname
  {{\t@ghead\advance\tagnumber by -2\relax\number\tagnumber}}
\expandafter\def\csname @qnnum-1\endcsname
  {{\t@ghead\advance\tagnumber by -1\relax\number\tagnumber}}
\expandafter\def\csname @qnnum0\endcsname
  {\t@ghead\number\tagnumber}
\expandafter\def\csname @qnnum+1\endcsname
  {{\t@ghead\advance\tagnumber by 1\relax\number\tagnumber}}
\expandafter\def\csname @qnnum+2\endcsname
  {{\t@ghead\advance\tagnumber by 2\relax\number\tagnumber}}
\expandafter\def\csname @qnnum+3\endcsname
  {{\t@ghead\advance\tagnumber by 3\relax\number\tagnumber}}

\def\equationfile{%
  \@qnfiletrue\immediate\openout\eqnfile=\jobname.eqn%
  \def\write@qn##1{\if@qnfile\immediate\write\eqnfile{##1}\fi}
  \def\writenew@qn##1{\if@qnfile\immediate\write\eqnfile
    {\noexpand\tag{##1} = (\t@ghead\number\tagnumber)}\fi}
}

\def\callall#1{\xdef#1##1{#1{\noexpand\call{##1}}}}
\def\call#1{\each@rg\callr@nge{#1}}

\def\each@rg#1#2{{\let\thecsname=#1\expandafter\first@rg#2,\end,}}
\def\first@rg#1,{\thecsname{#1}\apply@rg}
\def\apply@rg#1,{\ifx\end#1\let\next=\relax%
\else,\thecsname{#1}\let\next=\apply@rg\fi\next}

\def\callr@nge#1{\calldor@nge#1-\end-}
\def\callr@ngeat#1\end-{#1}
\def\calldor@nge#1-#2-{\ifx\end#2\@qneatspace#1 %
  \else\calll@@p{#1}{#2}\callr@ngeat\fi}
\def\calll@@p#1#2{\ifnum#1>#2{\@rrwrite{Equation range #1-#2\space is bad.}
\errhelp{If you call a series of equations by the notation M-N, then M and
N must be integers, and N must be greater than or equal to M.}}\else%
 {\count0=#1\count1=#2\advance\count1
by1\relax\expandafter\@qncall\the\count0,%
  \loop\advance\count0 by1\relax%
    \ifnum\count0<\count1,\expandafter\@qncall\the\count0,%
  \repeat}\fi}

\def\@qneatspace#1#2 {\@qncall#1#2,}
\def\@qncall#1,{\ifunc@lled{#1}{\def\next{#1}\ifx\next\empty\else
  \w@rnwrite{Equation number \noexpand\(>>#1<<) has not been defined yet.}
  >>#1<<\fi}\else\csname @qnnum#1\endcsname\fi}

\let\eqnono=\eqno
\def\eqno(#1){\tag#1}
\def\tag#1$${\eqnono(\displayt@g#1 )$$}

\def\aligntag#1\endaligntag
  $${\gdef\tag##1\\{&(##1 )\cr}\eqalignno{#1\\}$$
  \gdef\tag##1$${\eqnono(\displayt@g##1 )$$}}

\def\eqalignno#1{\displ@y \tabskip\centering
  \halign to\displaywidth{\hfil$\displaystyle{##}$\tabskip\z@skip
    &$\displaystyle{{}##}$\hfil\tabskip\centering
    &\llap{$\displayt@gpar##$}\tabskip\z@skip\crcr
    #1\crcr}}

\def\displayt@gpar(#1){(\displayt@g#1 )}

\def\displayt@g#1 {\rm\ifunc@lled{#1}\global\advance\tagnumber by1
        {\def\next{#1}\ifx\next\empty\else\expandafter
        \xdef\csname @qnnum#1\endcsname{\t@ghead\number\tagnumber}\fi}%
  \writenew@qn{#1}\t@ghead\number\tagnumber\else
        {\edef\next{\t@ghead\number\tagnumber}%
        \expandafter\ifx\csname @qnnum#1\endcsname\next\else
        \w@rnwrite{Equation \noexpand\tag{#1} is a duplicate number.}\fi}%
  \csname @qnnum#1\endcsname\fi}

\def\ifunc@lled#1{\expandafter\ifx\csname @qnnum#1\endcsname\relax}

\let\@qnend=\end\gdef\end{\if@qnfile
\immediate\write16{Equation numbers written on []\jobname.EQN.}\fi\@qnend}

\catcode`@=12
\\

\font\twelverm=amr10 scaled 1200    \font\twelvei=ammi10 scaled 1200
\font\twelvesy=amsy10 scaled 1200   \font\twelveex=amex10 scaled 1200
\font\twelvebf=ambx10 scaled 1200   \font\twelvesl=amsl10 scaled 1200
\font\twelvett=amtt10 scaled 1200   \font\twelveit=amti10 scaled 1200

\skewchar\twelvei='177   \skewchar\twelvesy='60


\def\twelvepoint{\normalbaselineskip=12.4pt plus 0.1pt minus 0.1pt
  \abovedisplayskip 12.4pt plus 3pt minus 9pt
  \belowdisplayskip 12.4pt plus 3pt minus 9pt
  \abovedisplayshortskip 0pt plus 3pt
  \belowdisplayshortskip 7.2pt plus 3pt minus 4pt
  \smallskipamount=3.6pt plus1.2pt minus1.2pt
  \medskipamount=7.2pt plus2.4pt minus2.4pt
  \bigskipamount=14.4pt plus4.8pt minus4.8pt
  \def\rm{\fam0\twelverm}          \def\it{\fam\itfam\twelveit}%
  \def\sl{\fam\slfam\twelvesl}     \def\bf{\fam\bffam\twelvebf}%
  \def\mit{\fam 1}                 \def\cal{\fam 2}%
  \def\tt{\twelvett}
  \textfont0=\twelverm   \scriptfont0=\tenrm   \scriptscriptfont0=\sevenrm
  \textfont1=\twelvei    \scriptfont1=\teni    \scriptscriptfont1=\seveni
  \textfont2=\twelvesy   \scriptfont2=\tensy   \scriptscriptfont2=\sevensy
  \textfont3=\twelveex   \scriptfont3=\twelveex  \scriptscriptfont3=\twelveex
  \textfont\itfam=\twelveit
  \textfont\slfam=\twelvesl
  \textfont\bffam=\twelvebf \scriptfont\bffam=\tenbf
  \scriptscriptfont\bffam=\sevenbf
  \normalbaselines\rm}



\def\beginlinemode{\endmode
  \begingroup\parskip=0pt \obeylines\def\\{\par}\def\endmode{\par\endgroup}}
\def\beginparmode{\endmode
  \begingroup \def\endmode{\par\endgroup}}
\let\endmode=\par
{\obeylines\gdef\
{}}
\def\singlespace{\baselineskip=\normalbaselineskip}

\def\oneandahalfspace{\baselineskip=\normalbaselineskip
  \multiply\baselineskip by 3 \divide\baselineskip by 2}
\def\doublespace{\baselineskip=\normalbaselineskip \multiply\baselineskip by 2}

\newcount\firstpageno
\firstpageno=2
\footline={\ifnum\pageno<\firstpageno{\hfil}\else{\hfil\twelverm\folio\hfil}\fi}
\def\toppageno{\global\footline={\hfil}\global\headline
  ={\ifnum\pageno<\firstpageno{\hfil}\else{\hfil\twelverm\folio\hfil}\fi}}
\let\rawfootnote=\footnote		
\def\footnote#1#2{{\rm\singlespace\parindent=0pt\parskip=0pt
  \rawfootnote{#1}{#2\hfill\vrule height 0pt depth 6pt width 0pt}}}
\def\raggedcenter{\leftskip=4em plus 12em \rightskip=\leftskip
  \parindent=0pt \parfillskip=0pt \spaceskip=.3333em \xspaceskip=.5em
  \pretolerance=9999 \tolerance=9999
  \hyphenpenalty=9999 \exhyphenpenalty=9999 }
\def\dateline{\rightline{\ifcase\month\or
  January\or February\or March\or April\or May\or June\or
  July\or August\or September\or October\or November\or December\fi
  \space\number\year}}
\def\today{\ifcase\month\or
  January\or February\or March\or April\or May\or June\or
  July\or August\or September\or October\or November\or December\fi
  \space\number\day, \number\year}
\def\received{\vskip 3pt plus 0.2fill
 \centerline{\sl (Received\space\ifcase\month\or
  January\or February\or March\or April\or May\or June\or
  July\or August\or September\or October\or November\or December\fi
  \qquad, \number\year)}}


\hsize=6.5truein
\hoffset=0truein
\vsize=8.9truein
\voffset=0truein
\parskip=\medskipamount
\def\\{\cr}
\twelvepoint		
\doublespace		
\overfullrule=0pt	




\def\title			
  {\null\vskip 3pt plus 0.2fill
   \beginlinemode \doublespace \raggedcenter \bf}

\def\author			
  {\vskip 3pt plus 0.2fill \beginlinemode
   \singlespace \raggedcenter}

\def\affil			
  {\vskip 3pt plus 0.1fill \beginlinemode
   \oneandahalfspace \raggedcenter \sl}

\def\abstract			
  {\vskip 3pt plus 0.3fill \beginparmode
   \doublespace ABSTRACT: }

\def\submit  			
	{\vskip 24pt \beginlinemode
	\noindent \rm Submitted to: \sl}

\def\endtopmatter		
  {\endpage			
   \body}

\def\body			
  {\beginparmode}		

\def\head#1{			
  \goodbreak\vskip 0.5truein	
  {\immediate\write16{#1}
   \raggedcenter \uppercase{#1}\par}
   \nobreak\vskip 0.25truein\nobreak}

\def\beneathrel#1\under#2{\mathrel{\mathop{#2}\limits_{#1}}}

\def\refto#1{$^{#1}$}		

\def\references			
  {\head{References}		
   \beginparmode
   \frenchspacing \parindent=0pt \leftskip=1truecm
   \parskip=8pt plus 3pt \everypar{\hangindent=\parindent}}

\gdef\refis#1{\item{#1.\ }}			

\gdef\journal#1, #2, #3, 1#4#5#6{		
    {\sl #1~}{\bf #2}, #3 (1#4#5#6)}		

\def\endreferences{\body}

\def\figurecaptions		
  {\endpage
   \beginparmode
   \head{Figure Captions}
}

\def\endpage			
  {\vfill\eject}

\def\endpaper			
  {\endmode\vfill\supereject}


\def\heading				
  {\vskip 0.5truein plus 0.1truein	
   \beginparmode \def\\{\par} \parskip=0pt \singlespace \raggedcenter}

\def\subheading				
  {\vskip 0.25truein plus 0.1truein	
   \beginlinemode \singlespace \parskip=0pt \def\\{\par}\raggedcenter}

\def\tag#1$${\eqno(#1)$$}

\def\align#1$${\eqalign{#1}$$}

\def\aligntag#1$${\gdef\tag##1\\{&(##1)\cr}\eqalignno{#1\\}$$
  \gdef\tag##1$${\eqno(##1)$$}}

\def\endaligntag{}

\def\overset#1\to#2{{\mathop{#2}^{#1}}}
\def\underset#1\to#2{{\mathop{#2}_{#1}}}


\def\ref#1{Ref.~#1}			
\def\Ref#1{Ref.~#1}			
\def\[#1]{[\cite{#1}]}
\def\cite#1{{#1}}
\def\(#1){(\call{#1})}
\def\call#1{{#1}}
\def\taghead#1{}
\def\frac#1#2{{#1 \over #2}}

\def\12{{1\over2}}

\def\sla{\raise.15ex\hbox{$/$}\kern-.57em}
\def\leaderfill{\leaders\hbox to 1em{\hss.\hss}\hfill}
\def\twiddle{\lower.9ex\rlap{$\kern-.1em\scriptstyle\sim$}}
\def\bigtwiddle{\lower1.ex\rlap{$\sim$}}
\def\gtwid{\mathrel{\raise.3ex\hbox{$>$\kern-.75em\lower1ex\hbox{$\sim$}}}}
\def\ltwid{\mathrel{\raise.3ex\hbox{$<$\kern-.75em\lower1ex\hbox{$\sim$}}}}
\def\square{\kern1pt\vbox{\hrule height 1.2pt\hbox{\vrule width 1.2pt\hskip 3pt
   \vbox{\vskip 6pt}\hskip 3pt\vrule width 0.6pt}\hrule height 0.6pt}\kern1pt}
\def\tdot#1{\mathord{\mathop{#1}\limits^{\kern2pt\ldots}}}

\def\pmb#1{\setbox0=\hbox{#1}%
  \kern-.025em\copy0\kern-\wd0
  \kern  .05em\copy0\kern-\wd0
  \kern-.025em\raise.0433em\box0 }

\\
\catcode`@=11
\newcount\r@fcount \r@fcount=0
\newcount\r@fcurr
\immediate\newwrite\reffile
\newif\ifr@ffile\r@ffilefalse
\def\w@rnwrite#1{\ifr@ffile\immediate\write\reffile{#1}\fi\message{#1}}

\def\writer@f#1>>{}
\def\referencefile{
  \r@ffiletrue\immediate\openout\reffile=\jobname.ref%
  \def\writer@f##1>>{\ifr@ffile\immediate\write\reffile%
    {\noexpand\refis{##1} = \csname r@fnum##1\endcsname = %
     \expandafter\expandafter\expandafter\strip@t\expandafter%
     \meaning\csname r@ftext\csname r@fnum##1\endcsname\endcsname}\fi}%
  \def\strip@t##1>>{}}

\def\citeall#1{\xdef#1##1{#1{\noexpand\cite{##1}}}}
\def\cite#1{\each@rg\citer@nge{#1}}	

\def\each@rg#1#2{{\let\thecsname=#1\expandafter\first@rg#2,\end,}}
\def\first@rg#1,{\thecsname{#1}\apply@rg}	
\def\apply@rg#1,{\ifx\end#1\let\next=\relax
\else,\thecsname{#1}\let\next=\apply@rg\fi\next}

\def\citer@nge#1{\citedor@nge#1-\end-}	
\def\citer@ngeat#1\end-{#1}
\def\citedor@nge#1-#2-{\ifx\end#2\r@featspace#1 
  \else\citel@@p{#1}{#2}\citer@ngeat\fi}	
\def\citel@@p#1#2{\ifnum#1>#2{\errmessage{Reference range #1-#2\space is bad.}
    \errhelp{If you cite a series of references by the notation M-N, then M and
    N must be integers, and N must be greater than or equal to M.}}\else%
 {\count0=#1\count1=#2\advance\count1
by1\relax\expandafter\r@fcite\the\count0,%
  \loop\advance\count0 by1\relax
    \ifnum\count0<\count1,\expandafter\r@fcite\the\count0,%
  \repeat}\fi}

\def\r@featspace#1#2 {\r@fcite#1#2,}	
\def\r@fcite#1,{\ifuncit@d{#1}		
    \expandafter\gdef\csname r@ftext\number\r@fcount\endcsname%
    {\message{Reference #1 to be supplied.}\writer@f#1>>#1 to be supplied.\par
     }\fi%
  \csname r@fnum#1\endcsname}

\def\ifuncit@d#1{\expandafter\ifx\csname r@fnum#1\endcsname\relax%
\global\advance\r@fcount by1%
\expandafter\xdef\csname r@fnum#1\endcsname{\number\r@fcount}}

\let\r@fis=\refis			
\def\refis#1#2#3\par{\ifuncit@d{#1}
    \w@rnwrite{Reference #1=\number\r@fcount\space is not cited up to now.}\fi%
  \expandafter\gdef\csname r@ftext\csname r@fnum#1\endcsname\endcsname%
  {\writer@f#1>>#2#3\par}}

\def\r@ferr{\endreferences\errmessage{I was expecting to see
\noexpand\endreferences before now;  I have inserted it here.}}
\let\r@ferences=\references
\def\references{\r@ferences\def\endmode{\r@ferr\par\endgroup}}

\let\endr@ferences=\endreferences
\def\endreferences{\r@fcurr=0
  {\loop\ifnum\r@fcurr<\r@fcount
    \advance\r@fcurr by 1\relax\expandafter\r@fis\expandafter{\number\r@fcurr}%
    \csname r@ftext\number\r@fcurr\endcsname%
  \repeat}\gdef\r@ferr{}\endr@ferences}


\let\r@fend=\endpaper\gdef\endpaper{\ifr@ffile
\immediate\write16{Cross References written on []\jobname.REF.}\fi\r@fend}

\catcode`@=12

\citeall\refto		
\citeall\ref		%
\citeall\Ref		%

\\